\documentclass[12pt,preprint]{aastex}
\def\paren#1{\left( #1 \right)} 

\def\bra#1{\left[ #1 \right]}
\def\angle#1{\left\langle #1 \right\rangle}
\begin{document}

\title{Gravitational Radiation from Gamma-Ray Burst Progenitors}
\author{Shiho Kobayashi and Peter M\'{e}sz\'{a}ros}
\affil{
Center for Gravitational Wave Physics, \\
Dept. of Astronomy \& Astrophysics and Dept. of Physics,\\
Pennsylvania State University, 104 Davey Lab, University Park, PA 16802}

%%%%%%%%%%%%%%%%%%%%%%%%%%%%%%%%%%%%%%%%%%%%%%%%%%%%%%%%

\begin{abstract}
We study gravitational radiation from various proposed gamma-ray burst 
(GRB) progenitor models, in particular compact mergers and massive
stellar collapses. These models have in common a high angular rotation
rate, and the final stage involves a rotating black hole and accretion
disk system. We consider the in-spiral, merger and ringing phases, and
for massive collapses we consider the possible effects of asymmetric 
collapse and break-up, as well  bar-mode instabilities in the disks. 
We  evaluate the order-of-magnitudes of the strain and frequency of
the gravitational waves expected from various progenitors, at distances
 based on occurrence rate estimates. Based on simplifying assumptions,
 we give estimates of  the probability of detection of gravitational
 waves by the advanced LIGO system from the different GRB scenarios.
\end{abstract}

%%%%%%%%%%%%%%%%%%%%%%%%%%%%%%%%%%%%%%%%%%%%%%%%%%%%%%%%%%%
%%%%%%%%%%%%%%%%%%%%%%%%%%%%%%%%%%%%%%%%%%%%%%%%%%%%%%%%%%%

\section{Introduction}
\label{sec:intro}
The recent discovery of long-lasting afterglows of gamma-ray bursts (GRBs) 
at longer wavelengths has revolutionized this field (e.g. van Paradijs, 
Kouveliotou \& Wijers 2000)  and has greatly increased confidence in the 
relativistic blast wave model as an explanation for the afterglow 
electromagnetic signatures. 
As of now, upward of forty GRBs have been accurately localized, 
belonging to the class of so-called long bursts ($\gamma$-ray durations 
upwards of 2 s), and found to be predominantly in active star-forming regions. 
Although we lack direct evidence about the nature of the central engine 
producing the relativistic flow, it is nevertheless widely accepted that 
GRBs are the results of catastrophic events involving either
compact stellar mergers or massive stellar collapses (for reviews, see
Piran 2000; M\'esz\'aros 2002; Woosley 2001).
Since the ratio of the total duration of the bursts to the variability 
timescales is large, GRB $\gamma$-rays must be produced by internal
shocks or other dissipative events. A key feature of internal shocks is 
that the observed gamma-ray variability reflects the variability in the 
activity of the central engine (Kobayashi, Piran \& Sari 1997). Since 
variability timescales as short as a millisecond are observed, the engine 
must contain a compact object of less than a few solar masses, otherwise 
the light crossing time becomes larger than the variability timescales.  

While the fastest variability timescale is comparable to the dynamical
timescale of a millisecond, burst durations are usually very much longer,
and the central engine must be active much longer than its dynamical time.
This suggests that GRBs are powered by accretion disks and that the 
accretion timescales determine the durations. The observed energy 
in GRBs requires a massive ($\ge 0.1 - 1 M_\odot$) disk. Such a massive
disk can form from the fall-back of debris during the formation of the 
compact object itself, which ultimately is likely to be a newborn black hole.  
Several scenarios could lead to a black hole-massive accretion disk system. 
This includes the merger of double neutron star binaries (Eichler et al 1989; 
Ruffert et al. 1997), neutron star - black hole binaries (Paczynski 1991; 
Janka et al. 1999), black hole - white dwarf binaries (Fryer et al. 1999a), 
black hole - helium star binaries (Fryer \& Woosley 1998; Zhang \& Fryer 2001) 

The present and foreseeable sensitivity of gravitational wave  detectors
is such that for likely sources, including GRB, the detections would be
difficult, and for this reason, much effort has been devoted to the 
development of data analysis techniques that can reach deep into the 
detector noise.  A coincidence between a gravitational wave signal and a 
gamma-ray signal would greatly enhance the statistical significance of the 
detection of the gravitational wave signal (Finn, Mohanty \& Romano, 1999;
Kochanek \& Piran 1993).  Therefore, it is of interest to study the 
gravitational wave emission from GRB associated with specific progenitors.
Another reason for doing this is that, since the $\gamma$-rays and the 
afterglow are thought to be produced at very large distances ($\ge 10^{13}$cm) 
from the central engine, we have only very indirect information about the 
nature of the latter. However, gravitational waves should be emitted from the 
immediate neighborhood of the GRB central engine itself, and their observation 
should give valuable information about its identity.

In the frequency band $\sim 10-1000$ Hz relevant for the laser 
interferometer gravitational wave observatory (LIGO), and other detectors 
such as VIRGO, GEO600 and TAMA300 which are currently in operation, being 
developed or planned, the most promising sources of gravitational radiation 
are thought to be coalescing compact binaries (e.g. Phinney, 1991; Ruffert 
et al 1997, Janka et al 1999).  Compact binary mergers may be responsible 
for short bursts (e.g. Woosley, 2001), but there is no observational 
supporting evidence for this as yet.  On the other hand, those GRB which 
have been unambiguously localized and identified (all of them long bursts 
with $t_\gamma \gtrsim $ 10 s) are likelier to be associated with 
massive stellar collapses (van Paradijs et al 2000). Numerical calculations
of gravitational wave radiation from massive rotating stellar collapses
have been done in the Newtonian approximation, for ``collapsar'' models of 
long GRB in 2D (e.g. Fryer et al. 1999a, McFadyen \& Woosley 1999) and
in 3D for general cases not intended as models for GRB (Rampp et al 1998).
Recently, Dimmelmeier et al. (2002) have performed relativistic
simulations of rotational supernova core collapse in axisymmetry.
These numerical estimates are not conclusive, as a number of effects 
(including general relativity, secular  evolution,  non-axisymmetric 
instabilities etc, see Rampp et al 1998) 
have been neglected, but they suggest that gravitational wave emission from 
massive collapses may be much less important than from compact binary mergers. 
On the other hand, recent semi-analytical estimates (Fryer, Holz \& Hughes 
2002; Davies et al 2002; van Putten 2001) have indicated that instabilities 
in the collapsing core or in the accretion disk of a collapsar GRB could lead 
to significantly stronger gravitational wave signals than expected 
from the previous numerical estimates.  It is therefore of interest to 
re-examine the gravitational wave signals expected from various  specific 
GRB progenitors that have been recently discussed, and based on  current 
astrophysical models, to consider the range of rates and strains expected
in each case, for comparison with the LIGO sensitivity. 

In \S 2 we review the mechanisms that lead to gravitational wave
emission from GRB. In \S 3 we apply them to each of the GRB progenitor models 
and estimate the expected rates,  strain and frequency of the gravitational
waves. In \S 4, the detectability of the gravitational waves with the advanced
LIGO system is discussed.  We give conclusions  in \S 5.   

\section{Emission Mechanisms}
\label{sec:emission}

The process of binary coalescence as a gravitational wave source is
in principle simpler to analyze than that of a massive stellar collapse, 
although they share some common features, especially in the later phases. 
The binary coalescence process can be divided into three phases: 
in-spiral, merger and ring-down (e.g., Flanagan \& Hughes 1998). 
(1) During the in-spiral phase, the gravitational radiation reaction time 
scale is much longer than the orbital period. As the binary loses energy by 
gravitational radiation, the masses gradually spiral in toward each other. 
(2) The merger begins when the orbital evolution is so rapid that
adiabatic evolution is not a good approximation, or when the masses
(if the radii are much larger than their gravitational radius, e.g. white
dwarfs and Helium stars), come into contact with each other. Then the two 
masses go through a violent dynamical  merger phase which leads to a black 
hole on a dynamical timescale, releasing a fraction of their rest mass 
energy in gravitational waves. However, a significant fraction of the stellar 
material retains too much angular momentum to cross the black hole horizon 
promptly. This creates a temporary disk of debris material around the BH, 
whose accretion over times long compared to a dynamic time can power a GRB jet. 
(3) The black hole, right after it forms, is initially deformed, and in a 
ring-down phase radiates away the energy associated with these deformations 
as gravitational waves, until it settles into a Kerr geometry.

Collapsars, i.e. massive stellar collapses leading to a GRB, require a high core 
rotation rate, which may be easier to achieve if the star is in a binary system, 
although this is not necessary (e.g. Woosley, 2001). The high rotation rate 
is required to form a centrifugally supported disk around a central, possibly
spinning black hole, to power a GRB jet. A high rotation rate, however, may be
conducive to the development of bar or fragmentation instabilities in the 
collapsing core or/and in the massive disk around the central object (Nakamura 
\& Fukugita 1989; Bonnell \& Pringle 1995; van Putten 2001, 2002; Davies et al. 
2002; Fryer, Holz \& Hughes 2002). The asymmetrically infalling matter also 
perturbs the black hole's geometry, which leads to rig-down gravitational 
radiation. The gravitational wave emission from collapsars can thus in principle 
be estimated in a similar way to what is done in binaries during the in-spiral, 
merger and ring-down phases, although with considerably larger uncertainties.

\subsection{In-spiral}
\label{sec:inspiral}

In-spiraling compact binaries can be described as two point 
particles with masses $m_1$ and $m_2$, whose orbital parameters
evolve secularly due to gravitational radiation. The radiation carries
away orbital binding energy, which leads to a faster orbiting and more
compact system. Though the amplitude of the gravitational wave itself
$h(t)$ increases as the system evolves, the frequency $f$ also rapidly
increases. As the result, the energy spectrum is a decreasing function
of $f$ (e.g., Misner, Thorne \& Wheeler 1973),
\begin{equation}
\frac{dE}{df}=\frac{(\pi G)^{2/3}}{3}\mathcal{M}^{5/3}f^{-1/3},
\end{equation}
where $\mathcal{M}=(m_1m_2)^{3/5}(m_1+m_2)^{-1/5}$ is the chirp mass.  
The characteristic gravitational wave amplitude is defined with the
Fourier transform of $h(t)$ as $h_c(f)=f|\tilde{h}(f)|$, and equal 
to $\sim \sqrt{N} h$ where $N=f^2(df/dt)^{-1}$ is the number of cycles
radiated while the frequency changes by an amount of order $f$. 
The characteristic amplitude at a detector at a distance $d$ is given by
a function of the energy spectrum.
\begin{eqnarray}
h_{c} &=& \frac{1}{\pi d}\sqrt{\frac{G}{10c^3}\frac{dE}{df}},
\label{eq:hc}\\
&\sim& 1.4\times10^{-21}
\paren{\frac{d}{10Mpc}}^{-1}\paren{\frac{\mathcal{M}}{M_\odot}}^{5/6}
\paren{\frac{f}{100Hz}}^{-1/6},
\label{eq:hcrev}
\end{eqnarray}
This formula is obtained from an rms average of the amplitudes over 
different possible orientations of the source and interferometer.
Since we used the standard definition of the signal-to-noise ratio 
defined by eq. (\ref{eq:rho2}) in section \ref{sec:detectability},
our characteristic amplitude is smaller by a factor of $2\sqrt{5}$ than
eq. (5.1) in Flanagan \& Hughes (1998). For a given energy spectrum of
gravitational waves and noise spectrum density, both formulae give the
same signal-to-noise ratio.

\subsection{Merger}
\label{sec:merger}

Late in its evolution, a binary system will undergo a transition 
from an adiabatic in-spiral induced by gravitational radiation
damping to an unstable plunge induced by strong spacetime curvature 
(merger phase). For double neutron star binaries, the in-spiral
signal contribution can be taken to end around the frequency
(Kidder, Will \& Wiseman 1993; Lai \& Wiseman 1996), 
\begin{equation}
f_i \sim 1000 \paren{\frac{M}{2.8 M_\odot}}^{-1} \mbox{Hz},
\label{eq:fi}
\end{equation}
where the total mass $M=m_1+m_2$. However, if a white dwarf or a helium
star consists of the binary, the masses collide with each other at a
separation $l$ comparable to the sizes of masses. Since this happens
well before the relativistic or tidal effect becomes important, the
in-spiral gravitational signal ends at a much lower frequency,
\begin{equation}
f_i \sim 0.1 \paren{\frac{M}{M_\odot}}^{1/2} 
\paren{\frac{l}{10^9\mbox{cm}}}^{-3/2} \mbox{Hz}.
\end{equation}

The gravitational wave emission resulting from the coalescence of either 
double neutron star or black hole -- neutron star binary systems is still
poorly understood. The gravitational fields are quite strong and dynamical, 
which rules out a perturbative approach, and requires solving the full Einstein 
equations. Recently, numerical simulations of the coalescence of two equal mass
black holes in off-axis collisions have been calculated and provide some guidance 
(Khanna et al. 1999). They find that about $1\%$ of the total mass energy will 
emerge as gravitational waves during the final stages of the collision (ring-down 
phase). The radiation from the early merger stage of coalescence could be very 
much larger than the late stage ring-down radiation. Therefore, we assume that 
total energy radiated in the merger phase is  
\begin{equation}
E_m = \epsilon_m \paren{\frac{4\mu}{M}}^2 Mc^2 
\end{equation}
where $\epsilon_m$ (of nominal value 5\%) is a parametrisation of the total mass 
energy radiated in the coalescence, and the reduced mass is $\mu=m_1m_2/M$. 
The reduction factor $(4\mu/M)^2$ is unity for equal masses and gives the correct 
scaling law in the test particle limit $\mu \ll M$. 

The frequency of gravitational radiation in the in-spiral phase is
well-defined as a function of time, and increases monotonically. All
the energy emitted in the in-spiral phase is at frequencies less than
$f_i$. We assume that the spectrum of radiation in the merger phase
is confined to the frequency regime $f >f_i$.  As we discuss below, 
we define the end of the merger phase to occur when the waveform can be 
described by the $l=m=2$ quasi-normal mode signal of a Kerr
black hole. The quasi-normal ringing frequency $f_q$ gives an
approximate upper-bound for the frequencies carrying substantial power
during the merger (Flanagan \& Hughes 1998). 
\begin{equation}
f_q\sim \frac{F(a)c^3}{2\pi G M} 
\sim 32 F(a)\paren{\frac{M}{M_\odot}}^{-1} \mbox{kHz}.
\label{eq:qnm}
\end{equation}
where $F(a)=1-0.63 (1-a)^{3/10}$ and $a$ is the dimensionless spin
parameter of the black hole (Echeverria 1989). Though the 
energy spectrum could have some features related to the dynamical
instabilities (Zhuge et al. 1994; Dimmelmeier et al. 2002), we assume
the simplest flat spectrum with the following amplitude,
\begin{equation}
\frac{dE}{df} = \frac{E_m}{f_q-f_i} 
\end{equation}
Using eq (\ref{eq:hc}) and an approximation $f_q-f_i \sim f_q$, the
characteristic gravitational wave amplitude is given by 
\begin{equation}
h_c\sim 2.7\times10^{-22} F(a)^{-1/2}
\paren{\frac{\epsilon_m}{0.05}}^{1/2}
\paren{\frac{4\mu}{M}}
\paren{\frac{d}{10Mpc}}^{-1}
\paren{\frac{M}{M_\odot}}.
\label{eq:mer}
\end{equation}
Here the $M$, $\mu$ and $\epsilon_m$ are those appropriate to the end
of the merger  phase.

If dynamical instabilities develop in the rotating core or in the rotating 
massive disk during the merger phase, the deformed core/disk could radiate 
strong gravitational waves in a narrow frequency band. The deformation may
be considered, in its simplest form, as either two blobs or a bar.
Using in either case a formula appropriate for a rotating bar (e.g. Fryer, 
Holz \& Hughes 2002), we can estimate the amplitude of the corresponding
gravitational wave emission. Considering a bar of mass $m$ and length
$2r$ which rotates with  
angular frequency $\omega$, the mean strain is given by  
\begin{equation}
h=\sqrt{\frac{32}{45}}\frac{G}{c^4}\frac{mr^2\omega^2}{d}.
\end{equation}
Assuming $\omega^2=Gm^\prime/r^3$, the 
characteristic amplitude $h_c\equiv\sqrt{N}h$ is 
\begin{equation}
h_c\sim 1.9 \times 10^{-21}
\paren{\frac{N}{10}}^{1/2}
\paren{\frac{m}{M_\odot}}
\paren{\frac{m^\prime}{M_\odot}}
\paren{\frac{d}{10Mpc}}^{-1}
\paren{\frac{r}{10^6\mbox{cm}}}^{-1}.
\label{eq:bar}
\end{equation}
where we assumed that the waves remain coherent for $N=10$ cycles
(this value is optimistic).
When we discuss the instability of a rotating core, $m$ and $m^\prime$
are assumed to be equal and their cannot exceed the mass of the core, 
while when we discuss the instability of an accretion disk rotating around 
a black hole, $m$ and $m^\prime$ are assumed to be the masses of the 
accretion disk and the black hole, respectively.

\subsection{Ring-down}
\label{sec:ringdown}

A deformed black hole should undergo damped vibrations which emit
gravitational radiation. The most slowly damped mode, which has
spherical harmonic indices $l=m=2$, will dominate over other
quasi-normal modes at late times. Since this mode may also be
preferentially excited in the presence of binary masses or fragmentation
of a massive disk, we here focus attention on the $l=m=2$ mode.

The spectrum is peaked at $f_q$ with a width given by
the inverse of the damping time $\Delta f \sim \tau^{-1}=
\pi f_q/Q(a)$ where $Q(a)=2(1-a)^{-9/20}$ (Echeverria 1989),
\begin{equation}
\frac{dE}{df}\sim\frac{E_r f^2}{4\pi^4f_q^2\tau^3}
\bra{\frac{1}{\bra{(f-f_q)^2+(2\pi \tau)^{-2}}^2}+
\frac{1}{\bra{(f+f_q)^2+(2\pi \tau)^{-2}}^2}},
\label{eq:spec}
\end{equation}
where $E_r= \epsilon_r (4\mu/M)^2Mc^2$ is the energy radiated during the 
ring-down phase, and we assume $\epsilon_r=0.01$ as a nominal parameter henceforth.
The characteristic gravitational wave amplitude at $f=f_q$ is given by 
\begin{equation}
h_c\sim 2.0\times10^{-21}
\paren{\frac{\epsilon_r}{0.01}}^{1/2}
\paren{\frac{Q/F}{14}}^{1/2}
\paren{\frac{d}{10Mpc}}^{-1}\paren{\frac{\mu}{M_\odot}}.
\label{eq:qnm2}
\end{equation}
The value of the spin $a$ of the final black hole depends on the initial  
parameters of the system, and this dependence is not well understood at
present. Since the black hole may typically have spun up to near maximal 
rotation by a massive accretion disk (Thorne 1974), we adopt $a=0.98$
which is the value assumed by Flanagan \& Hughes (1998).

\section{Progenitors}
\label{sec:progenitors}

In recent years the black hole accretion disk model for GRBs has received
much attention (Fryer, Woosley \& Hartmann 1999; Popham, Woosley \& Fryer 1999;
M\'{e}sz\'{a}ros 2000, 2002; Narayan, Piran \& Kumar 2001). Progenitors likely 
to lead to this accretion system include binary mergers and collapsars:
double neutron stars (DNS), black hole - neutron star (BH-NS), black
hole - white dwarf (BH-WD), black hole - helium star (BH-He), and 
fast-rotating massive stellar collapses. If the viscosity parameter of the
disks has a standard value of $\alpha=0.1$, DNS and BH-NS mergers can explain 
short GRBs with durations under a second, but they are unlikely to produce 
long GRBs with durations of tens or hundred of seconds. On the other hand, 
BH-WD and BH - He mergers and collapsars might produce long GRBs.

Recently Fryer et al. (1999b) and Belczynski, Bulik \& Rudak (2002) have
estimated the formation rate of these progenitors by using population
synthesis methods. The results of Fryer et al. (1999b) are summarized in
Table 1, where the standard values of the formation rates and the uncertainty
ranges are listed. Assuming the galaxy density $n_{glx}=0.02$ Mpc$^{-3}$, 
we can estimate the distance inside which an event is expected to happen 
within in a year from the formation rates $R$. 
\begin{equation}
d \sim 230 \paren{\frac{R}{\mbox{Myr}^{-1}\mbox{galaxy}^{-1}}}^{-1/3}
\paren{\frac{n_{glx}}{0.02 \mbox{Mpc}^{-3}}}^{-1/3}\mbox{Mpc}.
\end{equation}
The estimates on the formation rates by Belczynski et al. (2002a) are
consistent with the results of Fryer et al. (1999b) and within the
uncertainty range in table 1 in most of their models. Though some of 
their models predict higher formation rates by
a factor of a few than the upper limits in table 1, the uncertainty
range of the distances are similar because the distances are rather
insensitive to the rate $d\propto R^{-1/3}$.

\begin{center}
Table 1  

\begin{tabular}{lcccc} \hline
      & \multicolumn{2}{c}{Formation Rate} & \multicolumn{2}{c}{Distance} \\
      & \multicolumn{2}{c}{[$\mbox{Myr}^{-1}\mbox{galaxy}^{-1}$]} 
      & \multicolumn{2}{c}{[Mpc]} \\
      & Standard & Range &  Standard & Range \\ \hline
DNS  & 1.2  & 0.01-80  & 220 & 53-1100 \\ 
BH-NS(a)           & 2.6  & 0.001-50 & 170 & 62-2300\\ 
\hspace{1.15cm} (b)& 0.55 & 0.001-50 & 280 & 62-2300\\ 
BH-WD& 0.15 & 0.0001-1 & 430 & 230-4900\\ 
BH-He& 14   & 0.1-50   & 95  &  62-490\\ 
Collapsar & 630 & 10-1000 & 27 & 23-110 \\ \hline
\end{tabular}
\end{center}

\subsection{Double Neutron Stars}
\label{sec:dns}

Stars more massive than $\sim 8 M_\odot$ are thought to collapse to form
a neutron star remnant and a core-collapse supernovae, while stars more 
massive than $\sim 20M_\odot$ are thought to result in a black hole 
remnant (Fryer 1999). The standard scenario to form close double neutron
star (DNS) binaries  
begins with two massive stars with masses between $\sim 8 M_\odot$ and 
$\sim 20 M_\odot$. 
The more massive (primary) star in the binary evolves off the main sequence, 
and forms a neutron star accompanied by a supernova explosion. 
If the system remains bound after the supernova explosion of
the primary, it results in a binary composed of a neutron star and a
massive main sequence star. When the secondary, in turn, evolves off the
main sequence and expands, the neutron star enters the hydrogen envelope
of the secondary and begins to spiral toward the secondary's helium core.
The orbital energy released ejects the hydrogen envelope, forming a
neutron star - helium star binary. After the explosion of the helium
star as a supernova, there remains a DNS binary.  

However, recent calculations of neutron star accretion reveal that,
during this common envelope phase, the neutron star can accrete over 
$1M_\odot$, and collapse to form a black hole
(Chevalier 1996; Bethe \& Brown 1998; Fryer et al. 1999b).  Thus, the
standard scenario for DNS binaries may in fact form BH-NS binaries
(see, however, Belczynski, Kalogera \& Bulik 2002).
An alternative scenario (Brown 1995) is that  the initial binary system
consists of two massive stars with nearly equal mass (within $5\%$ 
difference). The secondary evolves off the main sequence before the
explosion of the primary as a neutron star. The two stars then enter a
common envelope phase and form double helium star binary. After the
hydrogen envelope is ejected the helium stars explode and collapse to
neutron stars. 

\begin{figure}
\plotone{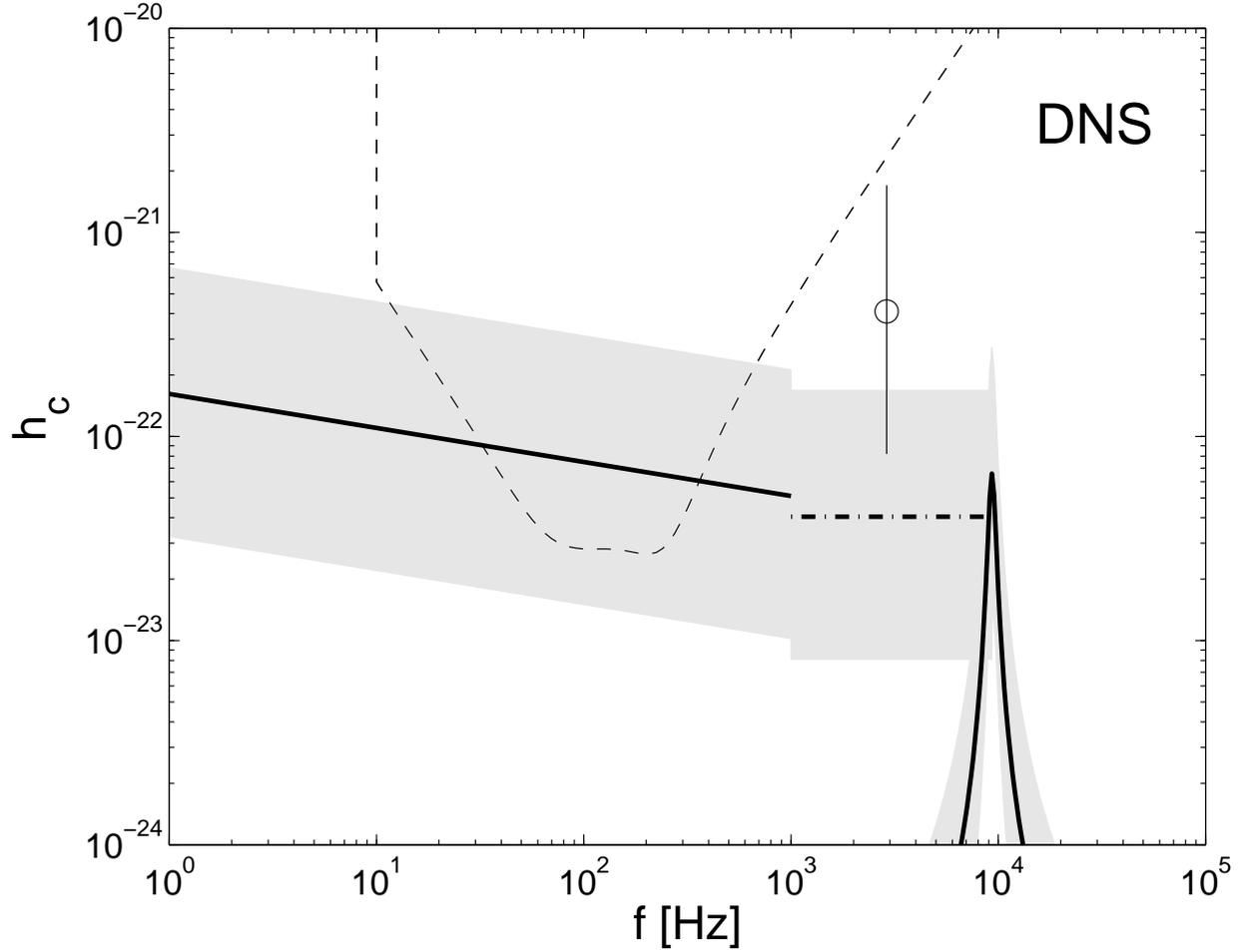}
\caption{
Double neutron stars: in-spiral (solid line), merger (dashed dotted line),
bar (circle), ring-down(solid spike), see discussion in \S
\ref{sec:dns}. $d=220$Mpc, $m_1=m_2=1.4M_\odot, a=0.98, \epsilon_m=0.05,
m=m^\prime=2.8M_\odot, N=10$ and $\epsilon_r=0.01.$.
Also shown is the advanced LIGO noise curve $\sqrt{f S_h(f)}$ (dashed curve). 
The shaded region and the vertical line reflect the uncertainty of the 
formation rate $R$ in Table 1. The values of $\epsilon_m$ and
$\epsilon_r$ are highly uncertain. The assumed values of
  $\epsilon_m=0.05$  and $\epsilon_m=0.01$ are rather optimistic. The
presented strain in the merger, bar and ring-down phases in Figure 1-5
give order-of-estimates or the upper limits.
 \label{fig:dns}}
 \end{figure}

The solid line in figure \ref{fig:dns} gives the characteristic 
amplitude of gravitational waves from a DNS binary 
($m_1=m_2=1.4M_\odot$) in the in-spiral phase at 220 Mpc.
This distance corresponds to the radius within which a merger event is
expected in a year, according to the rates in table 1. The shaded region
shows the uncertainty range corresponding to the formation rate
uncertainty. Around $f_i \sim 1000$ Hz, the orbital evolution becomes 
so rapid that adiabatic evolution is not a good approximation any more.
Then, the neutron stars begin to merge, they quickly form an object too
large to be supported by nuclear and degeneracy pressure. A black hole
forms on a dynamical time-scale, but a significant amount of mass $\sim
0.03-0.3 M_\odot$ will have, at first,  too much angular momentum to 
fall into the hole promptly (Fryer et al. 1999b). The resultant accretion 
disk and/or the spin of the black hole (via the Blandford-Znajek [1977]
mechanism) is responsible in this model for the relativistic jet which 
powers the electromagnetic signal of the GRB (M\'esz\'aros \& Rees 1997). 

The amplitudes of the gravitational wave signals associated with the 
merging (dashed-dotted) and ring-down phases (solid spike) of the black 
hole are also shown in Figure \ref{fig:dns}. The frequency of the BH 
quasi-normal mode is $\sim 9.3$kHz. In addition, a bar-mode instability 
might develop in the early stage of the merger, or a similar instability 
may arise intermittently in the inner disk. To illustrate the maximum signal 
levels that could arise from such instabilities, in equation (\ref{eq:bar}) 
we used arbitrary but plausible parameters $m=m^\prime=2.8M_\odot$, $r=4Gm/c^2$ 
and assuming that the waves remain coherent for $N=10$ cycles, we have  
plotted in the same figure the  characteristic strain (circle), with an 
error bar due to the uncertainty in the formation rate. 

\subsection{Black Hole -- Neutron Star }
\label{sec:bhns}

As discussed above, BH-NS binaries can form from neutron star
binaries in which the neutron star of the primary undergoes too much
accretion during the common envelope phase and collapses to a 
black hole. This formation scenario ($a$ in Table 1) produces a binary
consisting of a neutron star of $m_2 \sim 1.4M_\odot$  and 
a low-mass black hole of $m_1 \sim 3M_\odot$ (Fryer et al. 1999b).

The more standard formation scenario for BH-NS system (scenario $b$ in Table 1)
begins with two massive stars in which the primary have a mass greater
than  $\sim 20M_\odot$. The primary evolves off the main sequence, and
continues to evolve. When it forms a black hole, the system consists
of a black hole and a massive star. This system evolves through a common
envelope phase as the secondary star expands. During this common envelope
phase, the black hole spirals into the massive secondary and ejects the
hydrogen envelope. The supernova explosion of the secondary results in
the formation of a binary consisting of a neutron star and a higher mass 
black hole $m_1 \sim 12M_\odot$ (Fryer et al. 1999b).  Fryer et
al. (1999b) obtain the bimodal distribution of black hole masses
presented in their Fig 17, assuming that a black hole mass is equal to
1/3 of the progenitor mass. Though recent studies based on
core collapse simulations show that the mass distribution is rather flat
(Fryer \& Kalogera 2001; Belczynski et al. 2002b), we apply $m_1 \sim 12
M_\odot$ to the typical value of high mass black holes.

\begin{figure}
\plotone{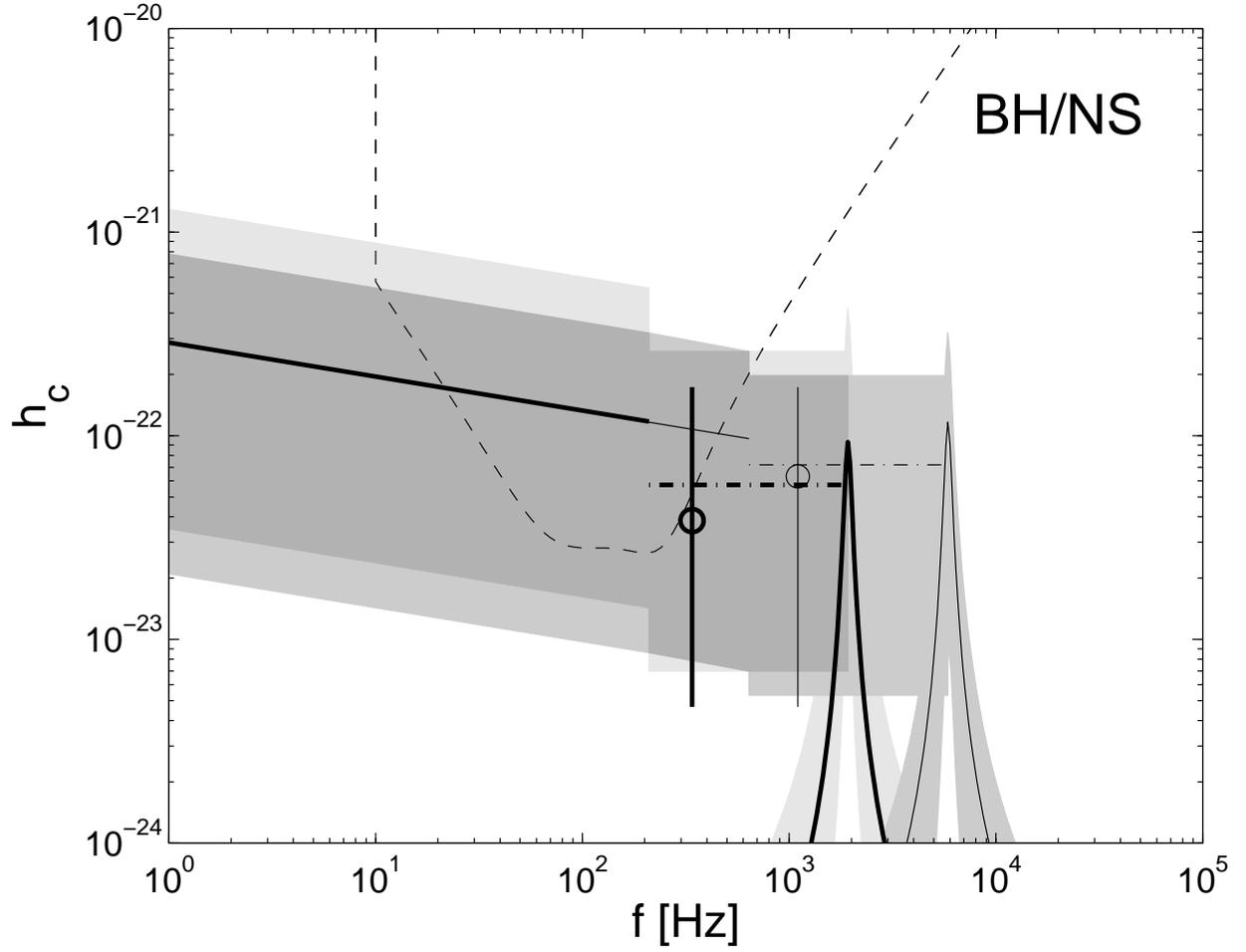}
\caption{
BH-NS, scenario a (thin lines and curve: $d=170$Mpc, 
$m_1=3M_\odot,m_2=1.4M_\odot,a=0.98,\epsilon_m=0.05,m=0.5M_\odot,
m^\prime=4M_\odot, N=10$ and $\epsilon_r=0.01$)
and scenario b (thick lines and
curve: $d=280$Mpc, 
$m_1=12M_\odot,m_2=1.4M_\odot,a=0.98,\epsilon_m=0.05,m=0.5M_\odot,
m^\prime=13M_\odot, N=10$ and $\epsilon_r=0.01$).
In-spiral (solid line), merger (dashed dotted line), bar (circle) and
  ring-down(solid spike).  
 \label{fig:bhns}} 
 \end{figure}

We assume that the gravitational wave  signal from the in-spiral binary
ends around the  frequency given by eq. (\ref{eq:fi}) and hence this signal 
shuts off nominally at $f_i \sim 640$Hz (scenario $a$) or $f_i\sim 210$Hz 
(scenario $b$). When the BH-NS binary merges, the neutron star is tidally 
disrupted. Some of the material accretes onto the BH directly, while the 
remainder forms an accretion disk of $0.3-0.7 M_\odot$ (Janka et al. 1999). 
The electromagnetic GRB signal from BH-NS is similar to that in DNS binaries,
and of comparable but somewhat larger energy (M\'esz\'aros \& Rees 1997).
The gravitational wave signals are also comparable, the differences
being associated with the BH mass (or possibly its spin rate, related to
the total accretion history). We show the amplitudes of the gravitational 
waves for the two BH-NS scenarios in figure \ref{fig:bhns}. The quasi-normal 
ring-down mode frequencies are $\sim 6$kHz ($a$) and $\sim 2$kHz ($b$). 
If a bar instability develops in the massive accretion disk, the maximum 
range of signals expected from this are calculated from equation (\ref{eq:bar})
with optimal but plausible parameters $m=0.5M_\odot$ and $r=6Gm^\prime/c^2$, 
rotating around a black hole of $m^\prime=4M_\odot$ ($a$) or $=13M_\odot$ ($b$), 
and assuming that the waves remain coherent for $N=10$ cycles. These are shown
as circles in figure \ref{fig:bhns}.

\subsection{Black Hole -- White Dwarf}
\label{sec:bhwd}

A scenario for binaries consisting of a black hole and a white
dwarf (BH-WD) begins with main sequence stellar systems having an extreme
mass ratio, with primary mass $\ge 20M_\odot$ and secondary mass
$\le 8 M_\odot$. After the primary collapses to form a black hole, 
one has a binary consisting of a black hole $\sim 3-15M_\odot$ and a 
main-sequence star (Fryer et al. 1999a). As the secondary star expands off 
the main sequence, a common envelope phase can occur, which results
in a decrease of the orbital separation. The system eventually evolves 
into a binary of a black hole and a high mass white dwarf of mass
$\ge 0.9M_\odot$ (Fryer et al. 1999a). Since white dwarf radii $\sim 10^9$cm
are much larger than those of neutron stars, the gravitational wave signal
from the in-spiral phase shuts off at a low frequency 
$f_i \sim 0.4 (M/10M_\odot)^{1/2}(r/10^9\mbox{cm})^{-3/2}$Hz.
This is well below the seismic noise cutoff $\sim 10$ Hz.  

\begin{figure}
\plotone{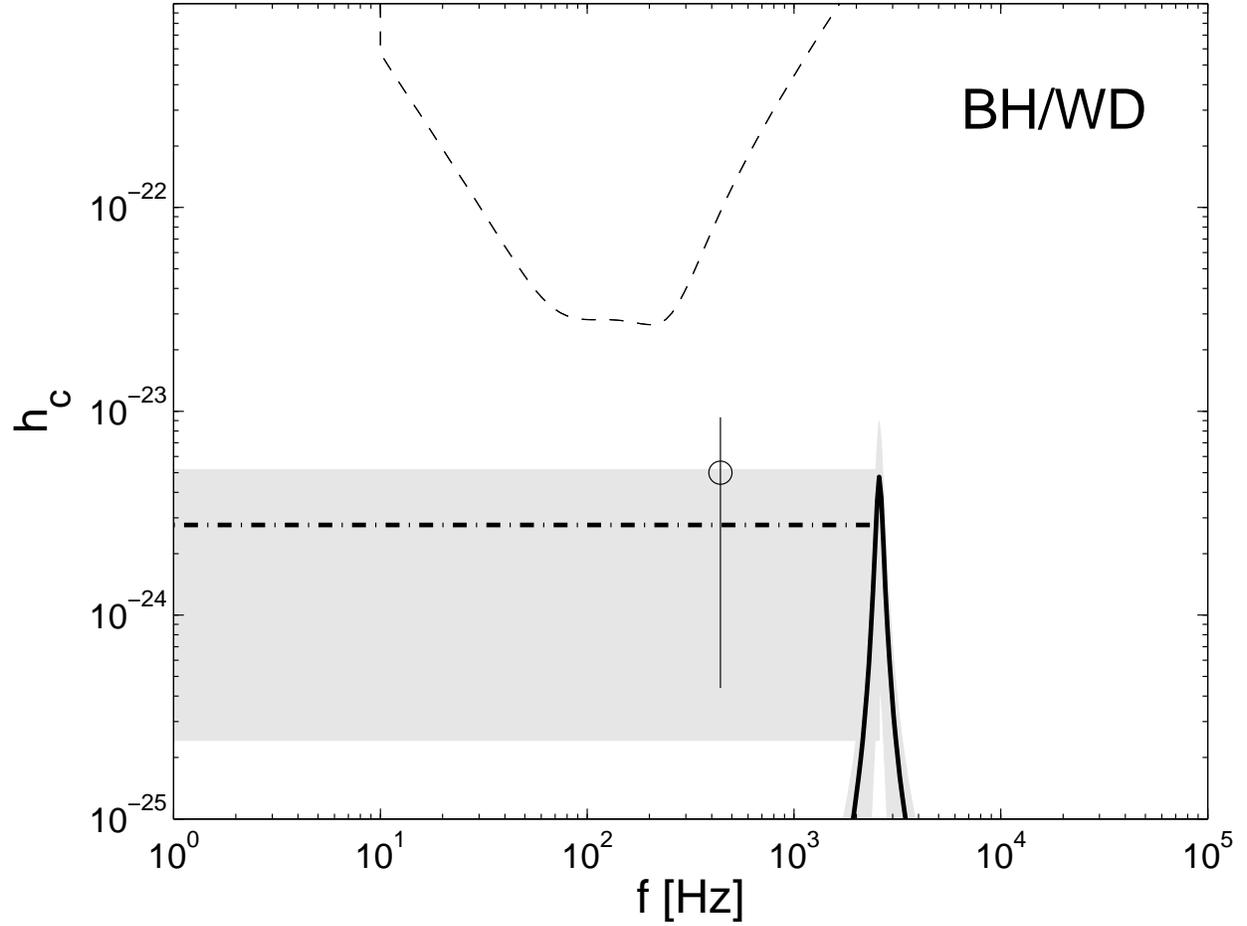}
\caption{
BH-WD: merger phase (dashed dotted line), bar (circle) and
ring-down(solid spike). $d=430$Mpc, 
$m_1=10M_\odot,m_2=0.1M_\odot,a=0.98,\epsilon_m=0.05,m=0.1M_\odot,
m^\prime=10M_\odot, N=10$ and $\epsilon_r=0.01$.
 \label{fig:bhwd}} 
 \end{figure}

As the BH-WD binary undergoes merger, the BH spirals inward through the
envelope of the white dwarf, in the process of which the white dwarf is 
tidally disrupted and most of its matter goes into an accretion disk around 
the black hole. Gravitational radiation is expected during this process
from the varying quadrupole moment of the BH and the inner portions of the
WD core or other inhomogeneities arising during the disruption. 
This can be described in a simplified manner as the gravitational radiation
of a BH and one or more discrete time-variable mass concentrations as they
approach and start to merge with each other. In figure \ref{fig:bhwd}, we give
the characteristic maximum gravitational wave amplitude from the coalescence 
of a BH of $m_1=10M_\odot$ and a white dwarf after the spiral-in phase
(with mass values motivated by Fryer et al 1999a), during the merger and 
disruption phase. The fraction of the WD mass concentration used in equation 
(\ref{eq:mer}) is $m_2= \alpha M_\odot$, where we take $\alpha \lesssim 10^{-1}$  
as a rough upper limit of the mass fraction of the WD core or blob, including 
any infall, which lies inside a radius $6 G m_1/c^2$ of the BH at the time of 
close approach to the BH when the dynamical merger process starts. 
Bar instabilities may also form in a disk resulting from the 
tidal destruction of the WD, and we have plotted the gravitational wave 
amplitude from a bar with $m= \beta M_\odot$ (with $\beta\sim 10^{-1}$) and 
length $r=6Gm^\prime/c^2$ rotating coherently for $N=10$ cycles around a black 
hole of $m^\prime= 10 M_\odot$, to represent the maximal radiation from 
bar-unstable massive disk. Also plotted is the final BH ring-down phase 
signal, which  is calculated from equation (\ref{eq:qnm2}) with $\mu$ 
estimated with the same masses as the final merger signal.

\subsection{Black Hole -- Helium Star}
\label{sec:bhhe}

A helium star (BH-He) merger model begins with a close binary system of
two massive stars, each greater than $8 M_\odot$. The more massive star 
(primary) evolves off the main sequence, and forms a compact object
(most likely a black hole) in a supernova explosion
(Fryer et al. 1999b; Belczynski et al. 2002a). When the secondary,
in turn, evolves off the main sequence and expands, the compact object
enters its companion's hydrogen envelope. If the in-spiraling compact 
object does not have sufficient orbital energy to eject the hydrogen
envelope, it moves on into the helium core. The compact object quickly
accretes enough material to become a black hole if it has not
already. The angular momentum of the black hole - helium core 
binary is injected into the helium core, forming a massive disk of mass
$\sim 4 M_\odot$ around a spinning black hole of $\sim 3 M_\odot$, with a
disk radius equal to a fraction of the initial helium core 
$\sim 10^{9}-10^{10}$ cm (Fryer \& Woosley 1998). 

\begin{figure}
\plotone{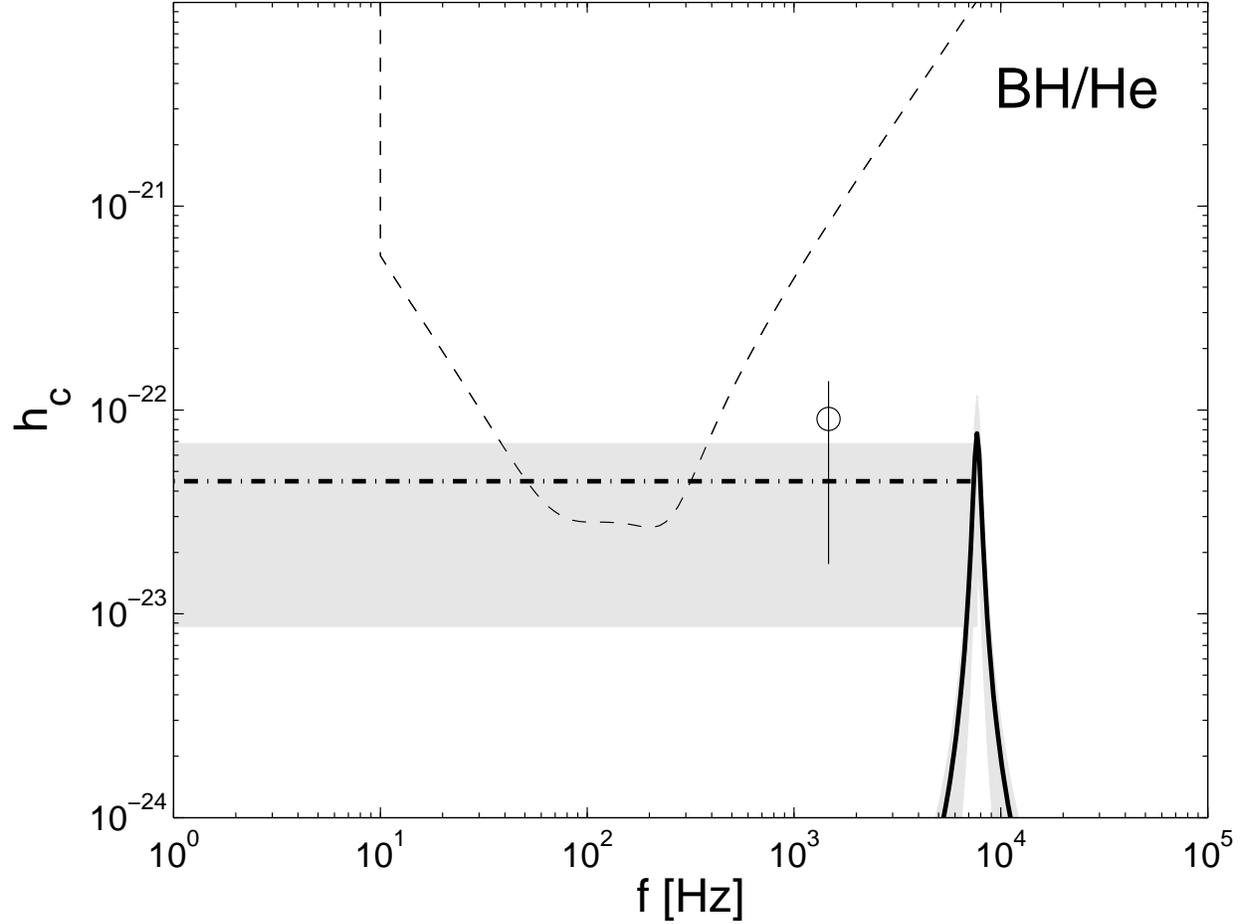}
\caption{
BH-He star: merger phase (dashed dotted line), bar (circle),
and ring-down(solid spike). $d=95$Mpc, 
$m_1=3M_\odot,m_2=0.4M_\odot,a=0.98,\epsilon_m=0.05,m=0.4M_\odot,
m^\prime=3M_\odot, N=10$ and $\epsilon_r=0.01$.
 \label{fig:bhhe}} 
 \end{figure}

A He star is of comparable or somewhat larger outer radius than a WD
and hence  gravitational waves from the pre-contact in-spiral phase again
lie below the seismic noise cut-off threshold. As in the BH-WD case, 
gravitational waves in the LIGO sensitivity range can arise after the
BH penetrates the envelope of the He star and circles its way inwards, due to
the varying quadrupole moment of the BH and the stellar core or inhomogeneities
arising during the tidal destruction of the star, from bar instabilities in the 
deformed massive tidal disruption disk, and from the final ring-down of the BH.  
In order to estimate the maximum characteristic amplitude in the merging 
and ring-down phases, we use eqs (\ref{eq:mer}) and (\ref{eq:qnm2}) with 
$m_1=3 M_\odot$ and $m_2=4 \alpha M_\odot$ (with mass values motivated
by Fryer \& Woosley 1998), where we took $\alpha\sim 10^{-1}$. 
The results are shown in figure \ref{fig:bhhe}. The maximum amplitude of 
the gravitational waves  from a bar with mass $m=4 \beta  M_\odot$ (where
$\beta\sim 10^{-1}$) and length $r=6Gm^\prime/c^2$ rotating coherently for 
$N=10$ cycles around a black hole with $m^\prime=3 M_\odot$ is also plotted.

\subsection{Collapsar}
\label{sec:coll}

If the progenitor mass exceeds $\sim 40M_\odot$, the collapse of
the iron  core does not produce a successful outgoing shock but instead
leads to the prompt formation of a black hole (type I collapsar). For a
GRB, it  is required that the progenitor, or at least its core, be fast
rotating  (MacFadyen \& Woosley 1999; Fryer 1999). As the inner stellar
layers fall  into the newborn black hole of $\sim 2-3 M_\odot$
\footnote{The final mass of the black hole depends on the
amount of fall-back, and could be as large as its progenitor (Fryer
1999). However, black hole - massive accretion disk systems in
the intermediate stage is relevant to GRB phenomena and the
gravitational radiation}
(MacFadyen \& Woosley 1999), and form
an accretion disk, the energy from  the massive disk and/or the rotation
of the black hole itself powers a GRB jet in the same way as in the
previous scenarios.

For progenitor masses between $\sim 20 M_\odot$ and $\sim 40 M_\odot$
(e.g. Fryer 1999) the collapse of the iron core leads to a different scenario, 
since the initial core mass infall is not sufficient to make a BH promptly, 
but leads to a temporary NS and an initial outward-moving shock.  After a 
delay of minutes to hours, the fall-back of stellar material that was initially 
moving outwards, but which fails to achieve escape velocity, finally drives
the central object above the maximal NS mass limit $\sim 3M_\odot$ (Rhoads \& 
Ruffini 1974), and a BH is formed, leading to a GRB. This is called the 
type II collapsar scenario. Although the accretion time-scales of type II 
collapsars may be too long for the average long bursts of $\sim 10-30$ sec, 
type II collapsars are probably as likely or somewhat more frequent events 
than Type I, because they involve a more densely populated portion of the 
stellar mass function. Assuming an initial mass function of $fdM \propto 
M^{-2.7}dM$ (Scalo 1986), the ratio of the number of progenitor stars of
both types is $N(20M_\odot-40M_\odot)/N(>40M_\odot) \sim 2.2$. 

Collapsars, even if arising from single stars, have in principle the potential 
to emit a substantial amount of their energy in gravitational waves, since 
they are fast-rotating and non-axisymmetries in the collapse are to expected. 
It is still a wide-open issue whether a rotating collapsing core 
fragments to produce two or more compact objects (Fryer et al. 2002). If it 
happens, their coalescence could emit strong gravitational waves
(Nakamura \& Fukugita 1989). Also, although the  mass of the accretion
disk in such systems depends on the viscosity assumed,  the disk mass is
as high as $\sim 1M_\odot$ in the low-viscosity models of MacFadyen \&
Woosley (1999). In this case, the self-gravity of the disk could become
important, and gravitational instabilities (e.g., spiral arm or bar  
formation) might develop and radiate gravitational waves (Davies et al 2002; 
Fryer et al 2002; van Putten 2002). The black hole will also suffer deformations 
from the non-uniform, non-axisymmetric accreting material, leading to ring-down.

\begin{figure}
\plotone{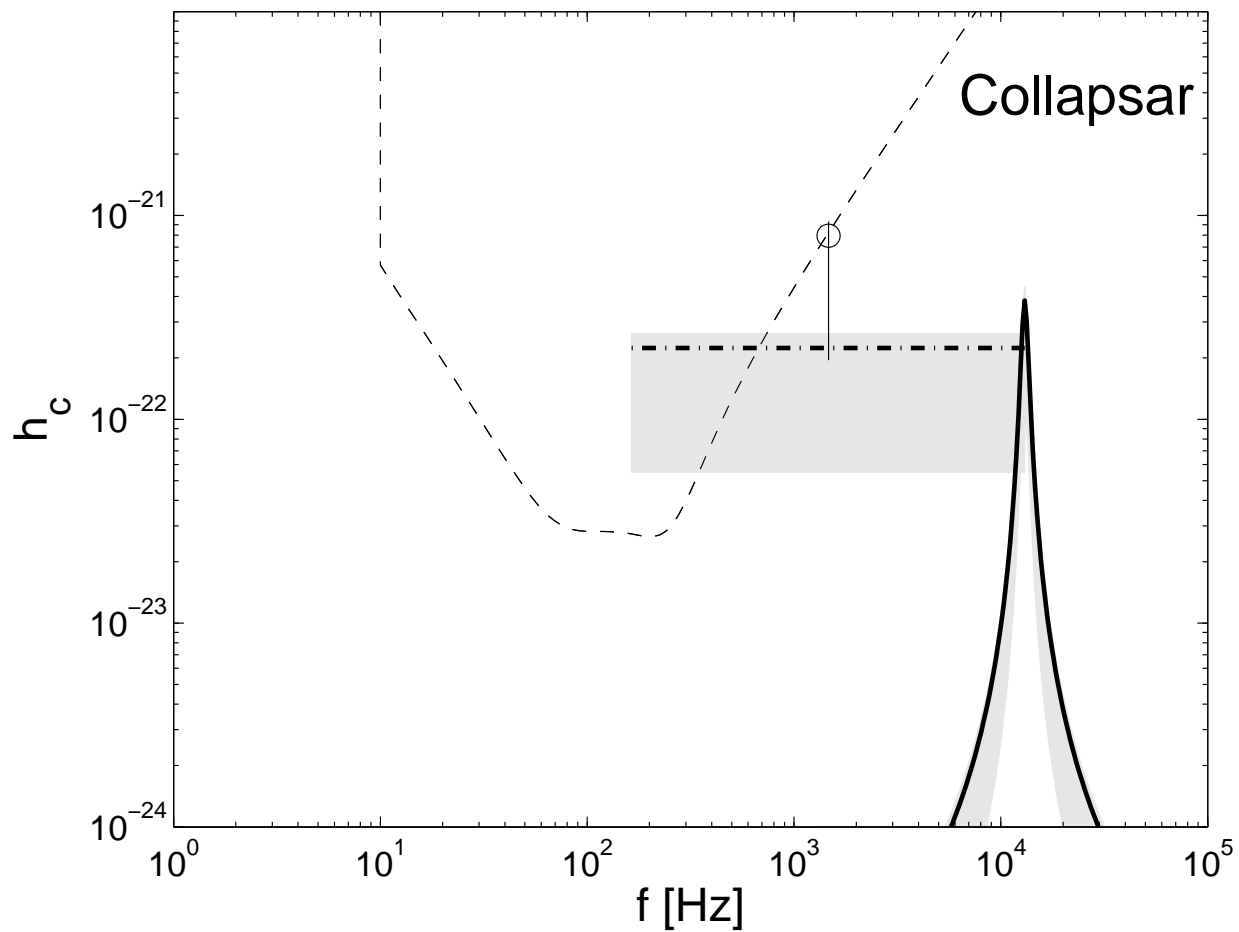}
\caption{
Collapsar: blob merger (dashed dotted line), bar (circle)
and ring-down(solid spike). $d=27$Mpc, 
$m_1=m_2=1M_\odot,a=0.98,\epsilon_m=0.05,m=1M_\odot,
m^\prime=3M_\odot, N=10$ and $\epsilon_r=0.01$, 
see discussion in \S \ref{sec:coll}.
 \label{fig:colla}} 
 \end{figure}

The characteristic gravitational wave amplitudes are calculated in a similar
way as before, under the assumption that the core collapse leads to
asymmetrical blobs which undergo a merger leading to a BH, which then
undergoes a ring-down phase.  The parameters chosen are plausible but 
arbitrarily chosen to represent the maximum level of signals that could 
be expected. These are shown in figure \ref{fig:colla}, where we assumed
$m_1=m_2= \alpha M_\odot$ with $\alpha \lesssim 1$ (representing blobs 
which become the BH) and we assumed that the merging 
phase starts when the blobs are separated by $r\sim10^7$ cm. 
A rough estimate on the maximal emission from an unstable accretion disk is 
given by the gravitational wave amplitudes from bars of 
$m=\beta M_\odot$, $m^\prime= 3 M_\odot$ and $r=6Gm^\prime/c^2$ 
where $\beta \lesssim 1$. 
We have calculated these amplitudes for the distances derived from the 
occurrence rates of Type I collapsars derived by Fryer et al. (1999b).
Type II rates were not computed, but if we take into account the Type II as 
well, using the mass function ratio in the next to last paragraph, the 
amplitude of gravitational waves from collapsars should be stronger than 
what is shown in figure \ref{fig:colla}, which corresponds to Type I only.

\section{Detectability}
\label{sec:detectability}

A complete set of theoretical waveform templates will be available for
the in-spiral and ring-down phases of compact binaries such as BH-BH,
and presumably also for BH-NS, NS-NS. 
These are directly applicable to
the first two GRB scenarios, and may be of some limited use in other
scenarios, e.g. if core collapse leads to break-up into very dense
(NS-like) blobs. When templates can be used, one can employ the matched 
filtering technique (e.g. Thorne 1987) to optimize the search for 
gravitational wave signals in the observational data stream. 
This technique is useful especially for in-spiral binaries emitting
signals of many cycles.
The signal-to-noise ratio (S/N) $\rho$ is given by  
\begin{equation}
\rho^2=4\int_0^\infty\frac{|\tilde{h}(f)|^2}{S_h(f)}df,
\label{eq:rho2}
\end{equation}
where $S_h(f)$ is the noise power spectral density of the detector. 
The gravitational wave signal is detectable if the S/N exceeds a threshold 
$\rho_{th}$, as a rough rule of thumb taken as $\rho_{th}\sim 5$.

{\it NS-NS and BH-NS --}
The matched filtering technique can in general be applied only to the 
in-spiral phases of NS-NS and BH-NS binaries.  Unfortunately, the ring-down 
frequencies of the quasi-normal modes of stellar mass black holes are too 
high to be optimal for the advanced LIGO. Furthermore, the in-spiral signal 
from BH-WD and BH-He binaries ends below the seismic cutoff frequency $\sim 
10$ Hz. 
For the nearest NS-NS binary in-spiral which happens in a year,
the S/N calculated numerically from equations (\ref{eq:hcrev}) and 
(\ref{eq:rho2}) is 
\begin{equation}
\rho_{NS-NS,insp} \sim 7.4 ~(1.5,~30.) ({\cal M}/1.2M_\odot)^{5/6}
(R/1.2 ~\mbox{Myr}^{-1}\mbox{galaxy}^{-1})^{1/3}~
\label{eq:rhonsns}
\end{equation}
where the numbers in parenthesis give the range of the S/N ratio due to
the uncertainty in the formation rate $R$ listed in Table 1, and ${\cal M}=
(m_1 m_2)^{3/5}(m_1 + m_2)^{-1/5}$ is the chirp mass.
For the nearest BH-NS binary in a year in scenario a and b, the S/N values and
their uncertainties are 
\begin{equation}
\rho_{BH-NS,insp}= \cases{ 13~(0.9,~35.) ({\cal M}/1.8 M_\odot)^{5/6}
     (R/2.6 ~\mbox{Myr}^{-1}\mbox{galaxy}^{-1})^{1/3}~,& scenario (a); \cr
                           12~(1.5,~54.) ({\cal M}/3.2M_\odot)^{5/6} 
      (R/0.55 ~\mbox{Myr}^{-1}\mbox{galaxy}^{-1})^{1/3}~,& scenario (b).\cr}
\label{eq:rhonsbh}
\end{equation}
The advanced LIGO system could therefore detect gravitational waves from 
these NS-NS and BH-NS scenarios.

{\it Collapsars --}
For collapsars, as well as for BH-WD and BH-He star binaries,
the gravitational wave frequencies in the merger phase (if the core breaks 
up into blobs) come into a suitable range for detection by the advanced LIGO. 
Although the waveform is unknown, the signals may nonetheless be detectable 
using the cross-correlation of the outputs of two LIGO detectors.
The two LIGO detectors are coaligned, but widely separated. If we set the 
arrival time of the GRB signal as the origin of time at each detector
by using the sky position of the GRB or the afterglow, we can correct for
the physical separation so that the two detectors can be considered
as ``coincident'' (i.e. having identical locations and arm orientations).  
The outputs of the two detectors around the onset of the GRB are   
\begin{equation}
s_1(t)=n_1(t)+h_1(t), \ \ \ s_2(t)=n_2(t)+h_2(t),
\end{equation}
where $h_1(t)$ and $h_2(t)$ are the gravitational wave strains 
in the two detectors, and $n_1(t)$ and $n_2(t)$ denote the noise 
components intrinsic to the two detectors. Since we assume that the 
two detectors are coincident and coaligned, the gravitational strains 
are essentially identical, $h(t)\equiv h_1(t)=h_2(t)$. Given the detector 
outputs, we can define the weighted cross-correlation signal $X_{on}$ as
\begin{equation}
X_{on}=\paren{s_1,s_2}=\int_{-T/2}^{T/2} dt \int_{-T/2}^{T/2}dt^\prime
s_1(t)s_2(t^\prime)Q(t-t^\prime),
\end{equation}
where $Q$ is a filter function. We can also write this equation in the
frequency domain.  
\begin{equation}
X_{on}\sim\int_{-\infty}^{\infty} df \int_{-\infty}^{\infty}df^\prime
\delta_T(f-f^\prime)\tilde{s}_1^*(f)\tilde{s}_2(f^\prime)
\tilde{Q}(f^\prime),
\end{equation}
where $\tilde{s}_1$, $\tilde{s}_2$ and $\tilde{Q}$ are the Fourier
transforms of $s_1(t)$, $s_2(t^\prime)$ and $Q(t-t^\prime)$.
$\delta_T(f)=\sin(\pi fT)/\pi f$ is the finite-time
approximation to the Dirac delta function. If we knew the signal $h(t)$,
we could construct a $Q$ that maximizes the S/N as 
$\tilde{Q}(f)=|\tilde{h}(f)|/S_1(|f|)S_2(|f|)$. $S_1(f)$ and $S_2(f)$
are the power spectral densities of the noises of the two
detectors. Finn et al. (1999) suggested to adopt this filter with
$|\tilde{h}(f)|^2$ assumed to be unity in the detector band if a
detailed knowledge of $h(t)$ is lacking. Therefore, assuming for simplicity
that the power spectral densities are identical, $S(f)\equiv S_1(f)=S_2(f)$
in the two detectors, we adopt $\tilde{Q}(f)=\lambda/S^2(|f|)$ where 
$\lambda$ is a normalization constant.

The expected value of the cross-correlation signal 
(averaged over the source population) is  
\begin{equation}
\angle{X_{on}}\sim \lambda \int_{-\infty}^{\infty} 
\frac{\angle{h_c^2(f)}}{f^2S^2(|f|)}df. 
\label{eq:Xon}
\end{equation}
Using data segments not associated with GRBs (off-source), we can
also evaluate the fluctuation of the cross-correlation of the noise,
\begin{equation}
\sigma^2_{off}=\overline{\paren{n_1,n_2}^2}
\sim \frac{\lambda^2 T}{4}\int_{-\infty}^{\infty} \frac{df}{S^2(|f|)} ~.
\label{eq:sigmaoff}
\end{equation}
The factor $T$ on the right hand side arises from evaluating 
$\delta_T(0)$. 
We assume that the duration of gravitational wave bursts are shorter 
than $T=10$ sec. In the previous section, we assumed that the merger 
phase starts at $\sim 200$ Hz (collapsars), or at $f_i\sim 0.4$Hz 
(BH-WD and BH-He binaries).  During $T=10$ sec, 
the system can rotate multiple times, and it is plausible to assume 
that a significant fraction of the energy in the system is emitted 
during this time. If the energy spectrum of the gravitational waves
$dE/df$ is flat, as we have assumed, most of gravitational wave energy 
is emitted in the high frequency band $\sim f_q$. The emission timescale 
might be comparable to the damping time of the quasi-normal mode $\tau$ 
(Flanagan \& Hughes 1998). Although $\tau$ is much shorter than 10 sec, 
the lag between a GRB signal and the gravitational wave signal are 
model-dependent and rather uncertain. We assume a conservative estimate
of $T=10$ sec. 

We define the S/N of the gravitational wave signal in the merger phase
as $\overline{X}_{on}/\sigma_{off}$. For the nearest  collapsar which 
occurs in a year, the S/N ratio calculated numerically from equations
(\ref{eq:Xon})(\ref{eq:sigmaoff}) is
\begin{equation}
\rho_{Coll, merg}\sim 3
\paren{\frac{\epsilon_m}{0.05}}
\paren{\frac{F(a)}{0.8}}^{-1}
\paren{\frac{T}{10 ~\mbox{sec}}}^{-1/2}
\paren{\frac{\mu}{0.5 M_\odot}}^2
\paren{\frac{R}{630 \mbox{Myr}^{-1}\mbox{galaxy}^{-1}}}^{2/3}
\label{eq:rhocoll}
\end{equation}
where the uncertainty range in $\rho$ from the uncertainty range in the 
rate $R$ is $0.2-3.8$ (from Table 1), the blob masses enter through the
reduced mass $\mu=m_1 m_2/(m_1+m_2)$, and we assumed that the blob merger
starts at $r=10^7$ cm (which determines the range of frequency overlap 
between the signal and noise curves of Figure \ref{fig:colla}), and 5\% of 
the gravitational energy of the merging blobs is radiated away.
For collapsars, this estimate is highly dependent on the assumed radius at 
which the merging phase of the blobs starts (and of course on the assumption 
that break-up into blobs is prevalent).  If we assume that the merger starts 
at $r \sim 3\times 10^7$ cm, instead of $10^7$ cm as assumed for the previous 
estimate above, the gravitational wave signal shown as a dot-dashed line
and shaded band in Figure \ref{fig:colla} moves further left, and completely 
overlaps the most sensitive band $\sim 50-500$Hz of the detector. The S/N 
is then $\rho_{Coll, merg}\sim 4.5~ (0.3-6.2)$, with other parameters 
scaling as in equation (\ref{eq:rhocoll}). 
The same equation (\ref{eq:rhocoll}) with a factor 4.5 in front is applicable 
also to BH-WD  mergers
(with $\mu \sim 0.1M_\odot$, $R\sim 0.15\mbox{Myr}^{-1}\mbox{galaxy}^{-1}$,
see \S \ref{sec:bhwd}) and BH-He  mergers
(with $\mu\sim 0.4M_\odot$, $R\sim 14\mbox{Myr}^{-1}\mbox{galaxy}^{-1}$,
see \S \ref{sec:bhhe}), with the signal frequency range also spanning the 
LIGO sensitivity curve. This leads to 
$\rho_{BH-WD,merg} \sim 7 \times10^{-4}$ ($5\times10^{-6}-2\times10^{-3}$),
and $\rho_{BH-He,merg} \sim 0.2$ ($7\times10^{-3}-0.4$), respectively.
Thus, for BH-He and BH-WD binaries the S/N ratio is too low. However,
for collapsars using the above parameters, the S/N is near the threshold 
of detectability. Therefore, gravitational waves from the  collapsar
scenario of GRB assuming break-up of the core into blobs could be marginally 
detectable through the cross-correlation technique. 

{\it BH-WD and BH-He --} 
The BH-WD and BH-He binary scenarios of GRB are the least favorable for 
detection, because the value of the gravitational wave cross-correlations 
between two detectors is very small. In this case, one might still 
obtain some information on a possible association between GRBs and 
gravitational wave bursts, or upper limits on the corresponding strains,
by collecting many samples.
If the gravitational wave bursts are associated with GRBs, the correlated 
output of the two detectors will be different during the times when GRBs occur
(on-source) than during other times not associated with a GRBs (off-source). 
A statistically significant difference between on- and off-source 
cross-correlation measures would support the association.  This difference 
can be measured using Student's $t-$test (Finn et al. 1999). 
If $\sqrt{N_{on}}\angle{X_{on}}/\sigma_{off}$ is greater than a critical
value (say 2.58 for 99$\%$ significance), where $N_{on}$ is the number of 
the on-source events, we can conclude that we have found evidence for the
association.  The number of BH-He/WD binary events needed to detect the 
association with $99\%$ confidence is 
\begin{equation}
N_{BH-He/WD} \sim \paren{\frac{\angle{X_{on}}/\sigma_{off}}{2.58}}^{-2}
\end{equation}
If we assume that GRBs within a Hubble radius ($\sim 3000$Mpc) are 
detectable and that the BH-He/WD event rate is uniform, 
the typical distance to the event is $\sim 2000$Mpc. The S/N of the 
gravitational wave signal from a binary at 2000Mpc is 
$\rho_{BH-WD,merg} \sim   3\times10^{-5}$ and
$\rho_{BH-He,merg} \sim   4\times10^{-4}$.
The required number of events is 
$N_{BH-WD} \sim 7\times 10^9 $
$(\epsilon_m/0.05)^{-2}$ 
$(F/0.8)^2$
$(T/10 ~\mbox{sec})$
$(\mu/0.1M_\odot)^{-4}$ 
$(d/2000Mpc)^{4}$ and
$N_{BH-He} \sim 4\times 10^7$ 
$(\epsilon_m/0.05)^{-2}$ 
$(F/0.8)^2$
$(T/10 ~\mbox{sec})$
$(\mu/0.4M_\odot)^{-4}$ 
$(d/2000Mpc)^{4}$.
We might be able to select only nearby events by
using redshift observations of the afterglows, if available. When we 
analyze the nearest $n$ events in a year, the typical distance $d$ is 
proportional to $n^{1/3}$. Since the number of events needed to detect 
the association is $\propto d^4 \propto n^{4/3}$, the number of years 
it takes to collect samples is $\propto n^{1/3}$. When we collect only
the nearest event in a year $n=1$, the number of events needed to detect  
the association is
$N_{BH-WD}\sim 10^7$ 
$(\epsilon_m/0.05)^{-2}$ 
$(F/0.8)^2$
$(T/10 ~\mbox{sec})$
$(\mu/0.1M_\odot)^{-4}$ 
$(R/0.15 ~\mbox{Myr}^{-1}\mbox{gal}^{-1})^{-4/3}$
and 
$N_{BH-He} \sim 200$
$(\epsilon_m/0.05)^{-2}$ 
$(F/0.8)^2$
$(T/10 ~\mbox{sec})$
$(\mu/0.4M_\odot)^{-4}$ 
$(R/14 ~\mbox{Myr}^{-1}\mbox{gal}^{-1})^{-4/3}$,
or equivalently, it takes $10^7$ years and 200 years to
collect the samples. The ranges of $N_{BH-WD}$ and $N_{BH-He}$ 
due to the uncertainty of the formation rates are
$10^6-10^{11}$ and $37-10^5$, respectively.
Thus, it is unlikely that one can detect an association between GRBs and 
gravitational waves from BH-WD or BH-He binaries.  However, 
with a relatively small number sample, it is still possible to set a
tight upper limit on the amplitude of the gravitational waves. If we 
have $N_{BH-WD/He}=30$ events of BH-WD or BH-He binaries 
at any distances and we
detect no signal of gravitational waves, we can give an upper limit on
the mean amplitude of the events $h_{c}< 7.2\times
10^{-23}(N_{BH-WD/He}/30)^{-1/4}(T/10~\mbox{sec})^{1/4}$. 

\section{Conclusions}
\label{sec:conclusions}

We have estimated the strains of gravitational waves from some of the 
most widely discussed current GRB progenitor stellar systems. 
If some fraction of GRBs are produced by double neutron star  or neutron star
-- black hole mergers, the gravitational wave chirp signal of the in-spiral
phase should be detectable by the advanced LIGO within one year, associated 
with the GRB electromagnetic signal. We have also estimated the signals from 
the black hole ring-down phase, as well as the possible contribution of a bar 
configuration from gravitational instability in the accretion disk following 
tidal disruption or infall in GRB scenarios. The assumed values of the
parameters related to the gravitational energy emitted during merging and 
ring-down phase may be very optimistic. Thus, our calculations may be
regarded as order-of-magnitude estimates for the upper-limits to the strains.
Among the other progenitor scenarios, 
the signals from black hole -- Helium star and black hole -- white dwarf merger 
GRB progenitors are the least likely to be detectable, due to the low
estimates obtained for the maximum non-axisymmetrical perturbations. 
For another possible type of GRB progenitor, the massive rotating stellar 
collapses or collapsars, the non-axisymmetrical perturbations may be stronger,
and the estimated formation rates are much higher than for other progenitors,
with typical distances correspondingly much nearer to Earth. This type of
progenitor is of special interest, since it has so far received the most 
observational support from GRB afterglow electromagnetic observations.
For collapsars, in the absence of detailed numerical 3D calculations 
specifically aimed at GRB progenitors, we have roughly estimated the 
strongest signals that might be expected in the case of bar instabilities
occurring in the accretion disk around the resulting black hole, and in the
maximal version of the recently proposed fragmentation scenario of the 
infalling core. Although the waveforms of the gravitational waves produced in 
the break-up, merger and/or bar instability phase of collapsars are not known, 
a cross-correlation technique can be used making use of two co-aligned detectors.
Under these assumptions, collapsar GRB models would be expected to be marginally 
detectable as gravitational wave sources by the advanced LIGO within one year 
of observations.

Collapsars are suspected to be responsible for many of the long GRBs, with 
$\gamma$-ray durations $t_\gamma \gtrsim 2$ s (van Paradijs et al 2000), 
which have an occurrence rate $\sim 2$Gpc$^{-3}$yr$^{-1}$, or $\sim 
0.1\mbox{Myr}^{-1}\mbox{galaxy}^{-1}$. 
The short GRBs, with $t_\gamma \lesssim 2$ s, have an occurrence rate 
which is somewhat uncertain, but there are indications that they might be 
ten times more frequent than long ones (Piran 2002). If compact binary mergers, 
such as double neutron stars and black hole -- neutron star mergers lead to GRB, 
these should be of the short variety (Popham et al 1999). Such mergers are 
relatively simpler systems, which are estimated with greater confidence to be
strong gravitational wave  emitters.  The actual occurrence rates would be 
higher by a factor of $2/\theta^2$ if the $\gamma$-ray emission of GRBs is 
not isotropic but is beamed within an opening angle $\theta$. Current afterglow 
observations (available so far mostly for long bursts of $t_\gamma \gtrsim 10$ s) 
indicate that long GRBs are beamed with opening angles of a few to a few ten 
degrees. There is so far no evidence for beaming in short GRB. However, binary 
mergers also have natural channels along their rotational axis (although not 
as constraining as the walls of the extended stellar cores in collapsars), and 
they may therefore also be beamed into some as yet undetermined opening angle
(which is thought to be wider than in long bursts).  The formation rates 
listed in Table 1 are consistent with the observed GRB rates when considering 
the beaming factors, if these are assumed similar for all progenitors. 
As a consequence of the beaming, GRBs may not be observable in $\gamma$-rays 
from all progenitor events, due to misalignment between the observer and 
jet axes. In such cases, the so-called ``orphan" afterglows in other 
electromagnetic wavebands (X-ray, optical etc), and/or other GRB related 
transient events such as X-ray flashes (e.g. van Paradijs et al 2000) may be 
detectable. Such orphan electromagnetic burst events would also be associated 
with the (much less collimated) gravitational wave signals, which vary only 
by a factor $\sim 2$ between the  equator and the pole, the latter being 
along the rotation or GRB jet axis.

We would like to thank L. Sam Finn, Ben Owen, Patrick Sutton and Ian
Jones for useful discussions and valuable comments. We acknowledge
support through the Center for Gravitational Wave Physics, which is 
funded by NSF under cooperative agreement PHY 01-14375, and through 
NSF AST0098416 and NASA NAG5-9192.

%%%%%%%%%%%%%%%%%%%%%%%%%%%%%%%%%%%%%%%%%%%%%%%%%%%%%%%%%%%%%%%
\newpage

\noindent {\bf References}\newline
Belczynski,K., Bulik,T. \& Rudak,B. 2002a, ApJ, 571, 394.\newline
Belczynski,K., Kalogera,V. \& Bulik,T. 2002b, ApJ, 572, 407.\newline
Bethe,H. \& Brown,G.E. 1998, ApJ, 506, 780.\newline
Blandford,R.D. \& Znajek,R.L. 1977, MNRAS, 179, 433.\newline
Brown,G.E. 1995, ApJ, 440, 270.\newline
Bonnell, I.A. \& Pringle, J.E. 1995, MNRAS, 273, L12.\newline
Chevalier,R.A. 1996, ApJ, 459, 322.\newline
Davies, M.B., King,A., Rosswong,S. \& Wynn,G. 2002, ApJ, 579, L63.\newline
Dimmelmeier,H., Font,J.A. \& Muller,E. 2002, A\&A, 393, 523.\newline
Eichler, D.,  Livio,M., Piran,T. \& Schramm,D.N. 1989, Nature, 340, 126.\newline
Echeverria, F. 1989, Phys. Rev. D, 40, 3194.\newline
Finn,L.S., Mohanty,S.D. \& Romano, J.D. 1999, Phys. Rev. D, 60, 121101.\newline
Flanagan,E.E. \& Hughes, S.A. 1998, Phys. Rev. D, 57, 4535.\newline
Fryer, C.L. 1999, ApJ, 522, 413.\newline
Fryer, C.L., Woosley,S.E., Herant,M. \& Davies,M.B. 1999a, ApJ, 520, 650.\newline
Fryer, C.L., Woosley,S.E. \& Hartmann,D.H. 1999b, ApJ, 526, 152.\newline
Fryer, C.L. \& Kalogera, V. 2001, ApJ, 554, 548.\newline
Fryer, C.L. \& Woosley, S.E. 1998, ApJ, 502, L9.\newline
Fryer, C.L., Holz, D.E. \& Hughes,S.A. 2002, ApJ, 565, 430.\newline
Janka,H.T., Thomas,E., Ruffert, M. \& Fryer,C.L. 1999, ApJ, 527, L39.\newline
Khanna, G. et al. 1999, Phys. Rev. Lett., 83, 3581.\newline
Kidder,L.E., Will, C.M. \& Wiseman,A.G. 1993, Phys. Rev. D, 47, 3281.\newline
Kobayashi,S., Piran,T. \& Sari, R 1997, ApJ, 490, 92.\newline
Kochanek,C. \& Piran, T. 1993, ApJ, 417, L17.\newline
Lai,D. \& Wisenman, A.G. 1996, Phys. Rev. D, 54, 3958.\newline
MacFadyen,A.I. \& Woosley,S.E. 1999, ApJ, 524, 262.\newline
M\'{e}sz\'{a}ros, P. and Rees, M.J. 1997, ApJ, 482, L29 \newline
M\'{e}sz\'{a}ros, P. 2000, Nucl. Phys. B, 80, 63.\newline
M\'{e}sz\'{a}ros, P. 2002, Annu. Rev. Astron. Astrophys., 40,137.\newline
Misner,C.W., Thorne,K.S. \& Wheeler,J.A., Gravitation (Freeman, 
San Francisco 1973).\newline
Nakamura,T. \& Fukugita,M. 1989, ApJ, 337, 466.\newline
Narayan,R., Piran,T. \& Kumar,P. 2001, ApJ, 557, 949.\newline
Paczynski,B. 1991, Acta Astronomica, 41, 257.\newline
Phinney,E.S. 1991, ApJ, 380, L17.\newline
Piran, T. 2000, Phys. Rep., 333, 529.\newline
Piran, T. 2002, Review talk given at GR16, preprint gr-qc/0205045.\newline
Popham,R., Woosley,S.E. \& Fryer,C. 1999, ApJ, 518,356.\newline
Rampp,M., Muller,E. \& Ruffert,M. 1998, A\&A, 332, 969.\newline
Rhoads,C.E. \& Ruffini,R 1974, Phys. Rev. Lett., 32, 324.\newline
Ruffert, M., Janka,H.T., Takahashi,K. \& Shaefer,G. 1997, A\&A, 319, 122.\newline
Scalo, J.M. 1986, Fundam. Cosmic Phys.,11,1.\newline
Thorne,K.S. 1974, ApJ, 191, 507.\newline
Thorne,K.S., in 300 Years of Gravitation, edited by S.W. Hawking
and W. Israel (Cambridge University Press, Cambridge, 1987) p 330-458.\newline
van Putten, M.H.P.M. 2001 , ApJ, 562, L51.\newline
van Putten, M.H.P.M. 2002 , ApJ, 575, L71.\newline
van Paradijs,J, Kouveliotou,C \& Wijers,R.A.M.J, 2000, 
Annu.Rev.Astron.Astrophys, 38, 379\newline
Woosley,S. 2001, Proc. Rome Symp., Eds. E. Costa, F. Frontera, and J. Hjorth
 (Springer:Berlin)  p. 257.\newline
Zhang,W. \& Fryer,C.L. 2001, ApJ, 550, 357.
Zhuge,X., Centrella,J.M. \& McMillan,L.W. 1994, Phys. Rev. D, 1994, 6247.\newline

%\end{document}

%%%%%%%%%%%%%%%%%%%%%%%%%%%%%%%%%%%%%%%%%%%%%%%%%%%%%%%
%%%%%%%%%%%%%%%%%%%%%%%%%%%%%%%%%%%%%%%%%%%%%%%%%%%%%%%
%%%%%%%%%%%%%%%%%%%%%%%%%%%%%%%%%%%%%%%%%%%%%%%%%%%%%%%
%%%%%%%%%%%%%%%%%%%%%%%%%%%%%%%%%%%%%%%%%%%%%%%%%%%%%%%
%%%%%%%%%%%%%%%%%%%%%%%%%%%%%%%%%%%%%%%%%%%%%%%%%%%%%%%
%%%%%%%%%%%%%%%%%%%%%%%%%%%%%%%%%%%%%%%%%%%%%%%%%%%%%%%
%%%%%%%%%%%%%%%%%%%%%%%%%%%%%%%%%%%%%%%%%%%%%%%%%%%%%%%
%%%%%%%%%%%%%%%%%%%%%%%%%%%%%%%%%%%%%%%%%%%%%%%%%%%%%%%
%%%%%%%%%%%%%%%%%%%%%%%%%%%%%%%%%%%%%%%%%%%%%%%%%%%%%%%
%%%%%%%%%%%%%%%%%%%%%%%%%%%%%%%%%%%%%%%%%%%%%%%%%%%%%%%
\end{document}